# Thermal and microclimatic behavior of OASIS schoolyard paving materials

Ghid KARAM[1,2], Maïlys CHANIAL[1,3], Arnaud GRADOS[4], Martin HENDEL[1,2*], Laurent ROYON[1]

[1] Université Paris Cité, LIED, UMR 8236, CNRS, F-75013, Paris, France

[2] Univ Gustave Eiffel, ESIEE Paris, département SEED, F-93162, Noisy-le-Grand, France

[3] Paris City Hall, Ecological Transition and Climate & Road and Transportation Divisions, F-75013, Paris, France

[4] Université Paris Cité, MSC, UMR 7057, CNRS, F-75013, Paris, France

*corresponding author: martin.hendel@u-paris.fr

## Abstract

As part of its Resilience Strategy, the City of Paris' OASIS program aims to contribute ot its adaptation to heatwaves and climate change by transforming schoolyards into climate shelters, namely via desealing and greening. In this context, a variety of alternative pavement materials have been proposed to replace the initial schoolyard pavement, composed of an asphalt sidewalk structure. In the context of the EU-funded ERDF UIA OASIS project, the thermal performance of these alternative materials and their impact in terms of urban cooling was explored. To this aim, five samples of reference and innovative schoolyard pavements were studied in the lab under heat-wave conditions. Alternative green, biosourced, recycled and reflective pavement solutions were compared to standard fine-aggregate asphalt concrete. Their performance was evaluated with regards to their contribution to the urban heat island phenomenon and to pedestrian heat stress, account for the typical use schedule of schoolyards. Green and biosourced materials were found to perform well for both indicators, while the standard and recycled solutions had poor UHI performance but had limited negative effects on daytime heat stress. The reflective pavement had better UHI performance but had high radiosity during daytime which can negatively affect pedestrian heat stress.

## Abbreviations

| | | | |
|---|---|---|---|
| $\alpha$ | albedo | l | latent heat of water vaporization, 2 260 kJ/kg |
| $\varepsilon$ | emissivity | | |
| $\sigma$ | Stefan-Boltzmann constant, $5,67.10^{-8}$ W/(m².K⁴) | L | incident LW radiation |
| | | LW | longwave radiation (3-100 μm) |
| E | evaporation rate, kg/s | OASIS | Openness, Adaptation, Sensitisation, Innovation and Social Ties |
| ERDF | European Regional Development Fund | | |
| FAA | Fine-aggregate asphalt | $R_n$ | net radiation, W/m² |
| H | convective heat flux, W/m² | S | incident SW radiation |
| h | convective transfer coefficient, W/(m².K) | SW | shortwave radiation (0,3-3 μm) |
| UHI | urban heat island | $T_{air}$ | air temperature, °C |
| J | radiosity, W/m² | $T_s$ | surface temperature, °C |
| | | UIA | Urban Innovative Action |

## Introduction

Paris and its surroundings are expected to witness an increase in both the frequency and intensity of heat waves by the year 2100 [1]. In urban environments, the consequences of these extreme weather events are amplified by the urban heat island (UHI) effect, namely on public health with increased mortality and co-morbidity [2]. In France, awareness of this challenge among decision-makers was raised following the 2003 heat-wave [3]. It has since been bolstered by several hot summers since 2019, with record temperatures and several days reaching and/or exceeding 40°C in many parts of the country.

In response, many cities have integrated climate change adaptation plans to their municipal strategies, which often include measures against extreme heat and to counter the UHI effect. These measures tend to address its principal mechanisms, i.e. radiative trapping and heat storage, lack of evapotranspiration, wind obstruction and anthropogenic heat release [4].

These plans typically include a wide variety of urban cooling techniques that may be deployed at different scales such as cool roofs or pavements, pavement-watering, urban vegetation, street trees, urban forests, shading or misting devices, etc. [5]–[11]. Among other measures, the City of Paris' OASIS program aims to transform the schoolyards of its 760 public pre-, elementary and middle schools by the year 2040 into a network of climate shelters. The program is one of several climate change adaptation measures found in its Resilience Strategy adopted in 2017.

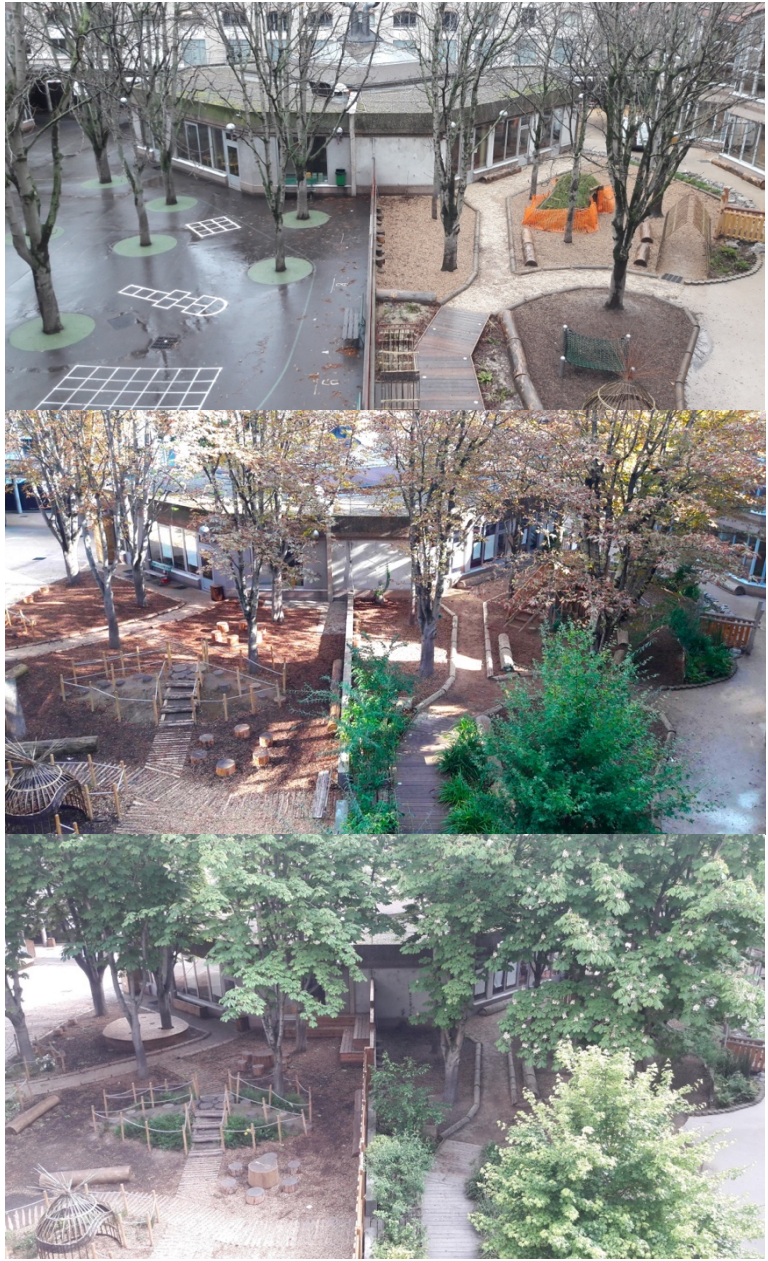

Figure 1 : Photographs of the Emeriau Elementary (left) and Kindergarten (right) OASIS schoolyards in February 2021 (top), September 2022 (middle) and May 2023 (bottom). In the kindergarten schoolyard, an artificial stream bordered with rocks runs from the right-hand side of the cafeteria (center building) downwards along the play area and under the wooden bridge and ends in the low-lying plants.

Traditional Parisian schoolyards generally have varying numbers of trees depending on the period the school was built. They generally offer very little low-lying vegetation and their surfaces are sealed, paved with an asphalt sidewalk structure composed of a fine-aggregate asphalt (FAA) laid on a concrete foundation. In response, the OASIS program promotes schoolyard unsealing and greening as well as the use of biosourced or recycled paving materials and children's games.

As part of the EU ERDF UIA OASIS project, 10 pilot schoolyards were monitored by a multidisciplinary scientific team, including an assessment of the microclimatic and thermal impacts of the schoolyard transformation works carried out during the summer of 2020. Figure 1 provides an illustration of such a transformation conducted at the Emeriau elementary school (left) and preschool (right) between 2021 and 2023. The Emeriau preschool was one of the 10 pilot schools evaluated during the ERDF project.

Among the evaluations that were conducted, the thermo-climatic behavior of five schoolyard pavement structures was studied in the laboratory under identical heatwave-like conditions. This was conducted using an existing experimental set-up used previously to study the thermo-climatic behavior of watered urban paving materials [12], [13]. A wide variety of cool pavement materials has been studied in the literature with the aim of reducing their contribution to UHI and to pedestrian heat stress [14]–[17]. To limit their contribution to UHI, cool pavements include modifications to lower their surface temperature and thus limit local air temperature heating via surface convection. While air temperature reductions contribute to reduce pedestrian heat stress, reflective pavements achieve this at the expense of higher radiosity, i.e. upwelling short- and longwave radiation, which may aggravate the pedestrian radiant heat load. While the resulting impact of these two counteracting effects on pedestrian heat stress depends on the considered time of day and on site characteristics, a growing number of studies have reported negative total effects of reflective pavements during daytime both experimentally and numerically [18]–[21]. It is therefore important to extend the analysis of pavement thermal behavior to include their convective and radiative contributions to the urban microclimate.

To this aim, the present article analyzes the thermal response of five OASIS schoolyard paving materials under heatwave conditions. The analysis aims to evaluate their thermo-climatic behavior with regards to their contribution to UHI and to pedestrian heat stress. The methodology deployed in previous experiments is improved upon to account for these counteracting effects.

## Materials and Methods

A brief description of the experimental setup and protocol is provided below. The interested reader may find a more detailed description in previous publications [22] and [23].

### Experimental Setup

Tested schoolyard pavement samples are subjected to a 24-hour cycle, divided into an 8-hour daytime period with artificial sunlight provided by a seven-bulb halogen lamp, and a 16-hour night-time period without sunlight.

During the day phase, the air temperature ($T_{air}$) is set to 35°C with a relative humidity of 35%; at night, they are set to 25°C and 70% respectively. The samples are first stabilized for 24 hours in night phase conditions before being subjected to the test cycle.

Type-T thermocouples and fluxmeters are positioned on the sample surface and at various depths to monitor its thermal behavior. A schematic diagram of the experimental set-up is shown in Figure 2. Table 1 summarizes the day and night phase setpoints for each climatic parameter controlled within the climate chamber.

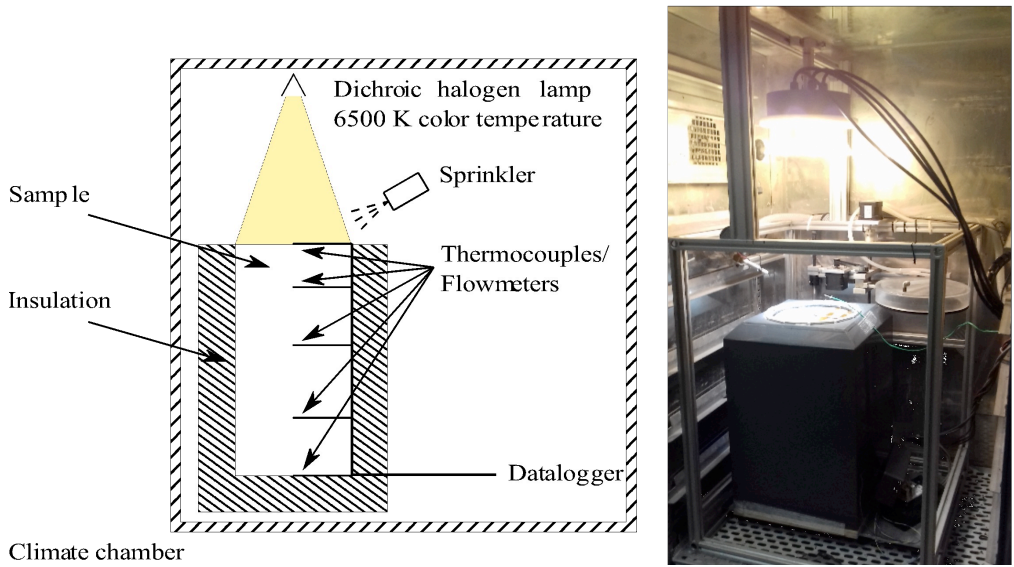

Figure 2: Experimental set-up diagram and photograph (from [22]).

Table 1: Day and night phase setpoint conditions

|  | Day | Night |
|---|---|---|
| **Duration** | 8h | 16h |
| **Air temperature** | 35°C | 25°C |
| **Relative humidity** | 35% | 70% |

## Tested Structures

The pavement structures studied are representative of those deployed within the OASIS pilot schoolyards before and after construction: conventional asphalt, light-colored asphalt, concrete pavers, wood chips and grass. The sample structures and sensor depths are illustrated in Figure 2.

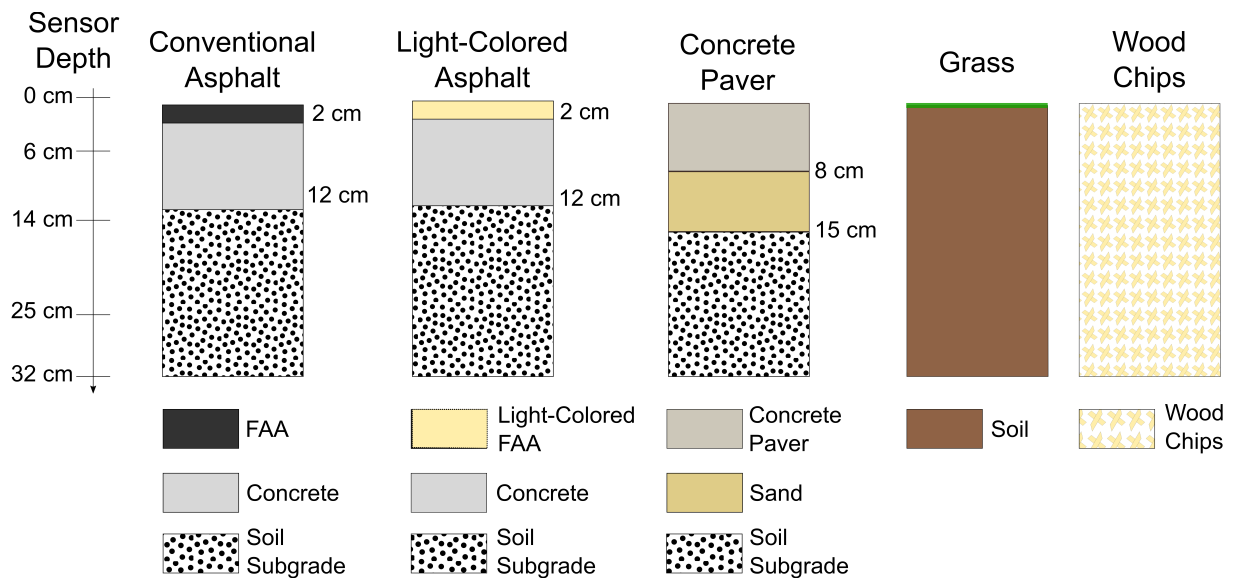

Figure 3: Studied sample structures and instrumentation depth (left axis)

The asphalt structures are composed of a 2 cm surface layer of fine-aggregate asphalt (FAA) on top of a 10 cm concrete foundation. This structure is typically found in Parisian sidewalks and has a smooth surface finish [13], [23]. While conventional FAA uses bitumen as a binder, a light-colored binder is used instead for the light-colored asphalt structure. As its name suggests, the concrete paver structure is composed of 8 cm thick concrete pavers laid on a 7 cm thick layer of sand. These materials are laid on the preexisting soil subgrade *in situ*. Grass and wood chip structures are composed of 32 cm of soil and wood chips, respectively.

Each cylindrical specimen is 32 cm in height and 16 cm in diameter and is insulated laterally and on the underside with a 5 cm-thick layer of polyurethane foam. The samples are monitored to simultaneously observe temperature and heat flow at different depths, except for the grass sample where only temperature is measured. Due to lamp ageing and replacements, SW and LW irradiance varies between experiments. Table 2 summarizes the average SW and LW irradiance for each sample.

Table 2 : Sample average SW and LW irradiance

| Sample | S (W/m$^2$) | L (W/m$^2$) |
|---|---|---|
| Conventional asphalt | 782 | 454 |
| Light-colored asphalt | 658 | 467 |
| Wood chips | 651 | 460 |
| Concrete pavers | 795 | 451 |
| Grass | 930 | 450 |

Material albedo was measured *in situ* in accordance with ASTM E1918-16 at the pilot schoolyards and in the laboratory according to ASTM E903-16. The first method uses an albedometer consisting of two pyranometers mounted back-to-back on a tripod-mounted arm placed 50 cm above ground level, pointing towards the sun. For the second method, an Agilent Cary 5000 UV-VIS-NIR spectrophotometer and 15-cm integrating sphere was used to characterize the spectral reflectivity of samples between 250 and 2500 nm. Reflectivity relative to the solar spectrum or to the solar lamp's spectrum is calculated by weighting the measured spectral reflectivity according to either source. *In situ* albedo as well as AM1.5 and halogen lamp albedo values are summarized in Table 3 for each structure.

Table 3: Sample albedo and emissivity

| Structure | Albedo | | | Emissivity |
|---|---|---|---|---|
| | Halogen lamp | AM1.5 | In situ | |
| Conventional asphalt | 0.06 | 0.06 | 0.14 | 0.98 |
| Light-colored asphalt | 0.40 | 0.33 | 0.34 | 0.98 |
| Wood chips[*] | 0.58 | 0.50 | 0.25 | 0.90 |
| Concrete pavers | 0.25 | 0.21 | 0.23 | 0.98 |
| Grass[*] | 0.25 | 0.25 | 0.17 | 0.95 |

---

[*] Grass and wood chip emissivity is taken from the literature [26], [27]

## Pavement Surface Energy Balance

Figure 4 illustrates the surface heat balance of an urban pavement.

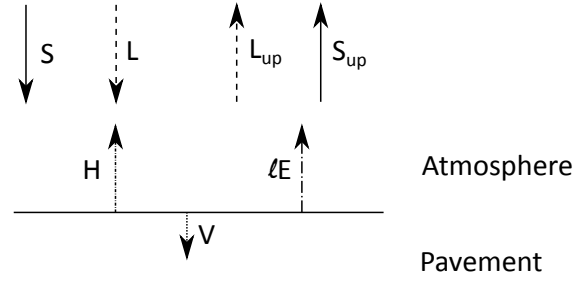

Figure 4: Pavement surface energy balance

$S$ and $L$ respectively correspond to SW and LW irradiance. $H$ is the atmospheric sensible heat flux (convection) and $V$ is the downwards conductive heat flux. Finally, $lE$ represents the latent heat flux due to surface water evapo(transpi)ration. $R_n$ is net radiation, i.e. the balance between incident irradiance ($S+L$) and radiosity ($S_{up}+L_{up}$). The surface energy balance equation is typically written:

$$R_n = H + lE + V \tag{1}$$

This equation can be rearranged to emphasize the inbound flows ($S + L$) on the one hand, and the outbound flows on the other:

$$S + L = S_{up} + L_{up} + H + lE + V \tag{2}$$

In the experiments described below, it should be noted that the latent heat flux is assumed to be zero for all tested samples except for the grass sample, which was thoroughly watered beforehand.

To limit the impact of a pavement on urban heating, attempts are generally made to reduce atmospheric convection $H$ between the pavement and the atmosphere. This parameter is directly proportional to the difference between the surface and air temperatures. Surface temperature is therefore a parameter that is targeted for reduction. In addition, a pavement's surface radiosity ($J = S_{up} + L_{up}$) has an impact on pedestrian heat stress by contributing to their radiant heat load, specifically during the day. In daytime, this term depends on both the surface temperature and the surface albedo of a given pavement. This is explained by detailing the expression of the terms $S_{up}$, $L_{up}$ and $H$ below:

$$S_{up} = \alpha S \tag{3}$$

with α the pavement's solar reflectivity, or albedo. This is a dimensionless number between 0 and 1, describing the surface's ability to reflect solar radiation.

$$L_{up} = (1 - \epsilon)L + \epsilon \sigma T_S^4 \tag{4}$$

with ϵ pavement emissivity, also a dimensionless number between 0 and 1, σ the Stefan-Boltzmann constant (5,67.10⁻⁸ W/m²/K⁴) and $T_S$ the pavement surface temperature. Emissivity and albedo are intrinsic optical properties of a given pavement surface and hence depend on the selected material.

A pavement's radiosity ($S_{up} + L_{up}$) therefore depends on its albedo, its emissivity and its surface temperature. Apart from untreated metals, most urban materials' emissivity is greater than 0.8. While some margin for improvement is available regarding emissivity, the principal parameter that can be used to reduce $T_S$ is albedo α.

Atmospheric convection $H$ depends on the surface-air temperature difference as well as the convective heat transfer coefficient $h$ as follows:

$$H = h\,(T_S - T_{air}) \tag{5}$$

The convective heat transfer coefficient is affected by site morphology and wind conditions. It is constant within the experimental setup, unchanging from one test to another [22]. More generally, it does not significantly depend on the pavement's properties, even though under natural convection (low-wind) conditions it will depend on the surface-air temperature difference to a certain extent.

Among these terms, convection contributes directly to heating the ambient air, impacting both pedestrian heat stress and UHI, particularly at night. Radiosity has a strong impact on pedestrian heat stress, particularly during the day in sunny areas. Both of these parameters will be analyzed hereafter to assess the impact of the studied pavement structures.

In order to eliminate the impact of variations in irradiance due to potential differences in experiment duration and/or the ageing and replacement of lightbulbs, we will compare the pavements in terms of their cumulative convection and radiosity, normalized by their cumulative irradiance. These quantities are described below:

$$H^* = \int \frac{H}{S+L} \tag{6}$$

$$J^* = \int \frac{S_{up}+L_{up}}{S+L} \tag{7}$$

These terms will be referred to as the pavements' convective and radiative contributions to the microclimate, with nighttime convective contributions of particular relevance for UHI mitigation and daytime radiative contributions of particular relevance for pedestrian heat stress for sun-exposed areas.

Although we won't focus our attention on them hereafter, analogous indicators for normalized conduction and evapo(transpi)ration can be defined as:

$$V^* = \int \frac{V}{S+L} \tag{8}$$

$$lE^* = \int \frac{lE}{S+L} \tag{9}$$

$V^*$ is representative of heat storage during the experiment cycle for a given sample. Assuming that the tested samples are thermally stabilized before the experiment, energy conservation can be expressed as:

$$H^* + J^* + V^* + lE^* = 1 \tag{10}$$

## Results

Experimental results are presented below, starting with temperature measurements, heat flux densities and finally microclimatic contributions of the studied pavement samples.

### Temperatures

#### Surface

Obtained surface temperature measurements are shown in Figure 6.

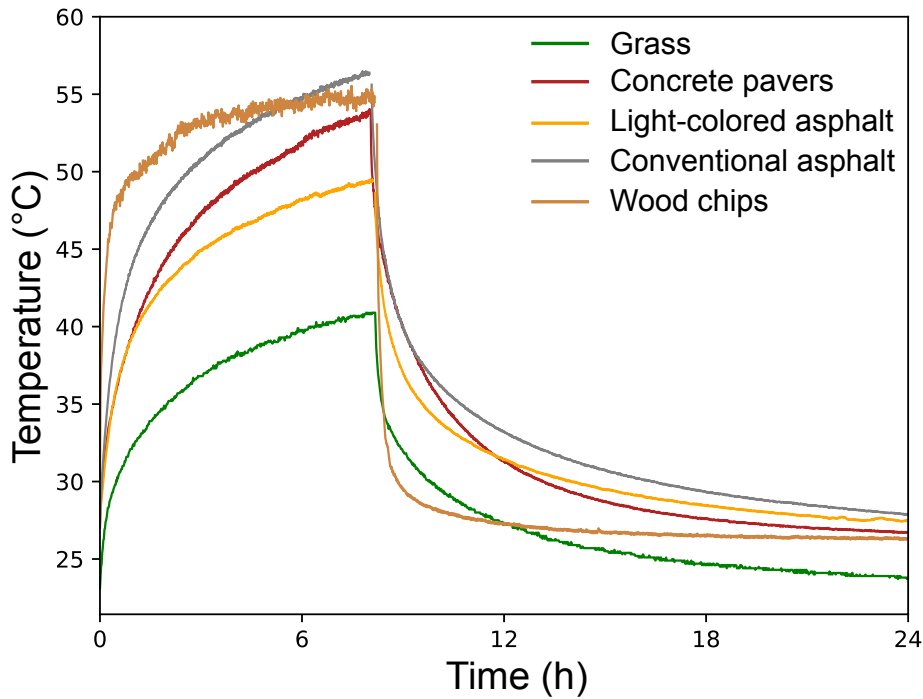

Figure 5: Surface temperature trends for different materials

The pavement structures heat up gradually during the daytime phase, reaching their maximum temperature after 8 hours, rising from 25°C to values between 40°C and 57°C. The night phase then begins, with the halogen lamp switched off and the set temperature reduced to 25°C. Surface temperatures first decrease suddenly, then more or less slowly, depending on the heat stored by the samples, converging toward the 25°C setpoint.

We can classify the various pavements according to their maximum surface temperature, with hot materials such as conventional asphalt reaching a surface temperature of 57°C, closely followed by wood chips and light-colored asphalt, whose maximum temperature approaches 55°C at the end of the daytime phase. Concrete pavers reach the same surface temperature as wood chips at the end of the daytime phase, having a comparable albedo, but take much longer to do so and do not stabilize by the end of the day phase. This testifies to the sample's high thermal admittance, which enables it to absorb the incident heat during the day and compensates for its significantly lower albedo compared to woodchips (0.25 vs. 0.58).

The light-colored asphalt is moderately warm during the day, with a maximum temperature of 47°C, while it stabilizes at the end of the night phase at a higher temperature than the concrete pavers. This is due partly to its higher albedo and partly to the high inertia of its concrete midcourse.

The studied lawn sample was saturated with water before the start of the experiment. As a result, it has the lowest surface temperature, with a maximum of around 40°C, despite receiving higher irradiance than the other samples. This is partly due to the evapotranspiration of the grass and soil, and partly to the high inertia of the waterlogged soil.

While the heating dynamics of the materials are characteristic of a first-order response, the response time for wood chips is significantly shorter than for the other materials, with a surface temperature that stabilizes at around 55°C only 4-5h into the day phase. On the other hand, the light-colored asphalt has the slowest response time among the dry samples.

At night, the surface temperature of all the structures decreases until it approaches the setpoint temperature. However, the cooling rate of the wood chips is significantly higher than that of the other materials, with an almost instantaneous drop in surface temperature when the lamp is switched off. This confirms its low heat accumulation during the day. Conversely, mineral materials, and in particular the light-colored asphalt among these, and grass have higher thermal inertia, slowly releasing the heat accumulated during the day.

**In Depth**

The samples are instrumented to observe temperatures 6 cm, 14 cm and 25 cm below the surface. Figure 7 illustrates temperature measurements at different depths. Temperature differences between depths of the same sample are indicative of heat propagation dynamics.

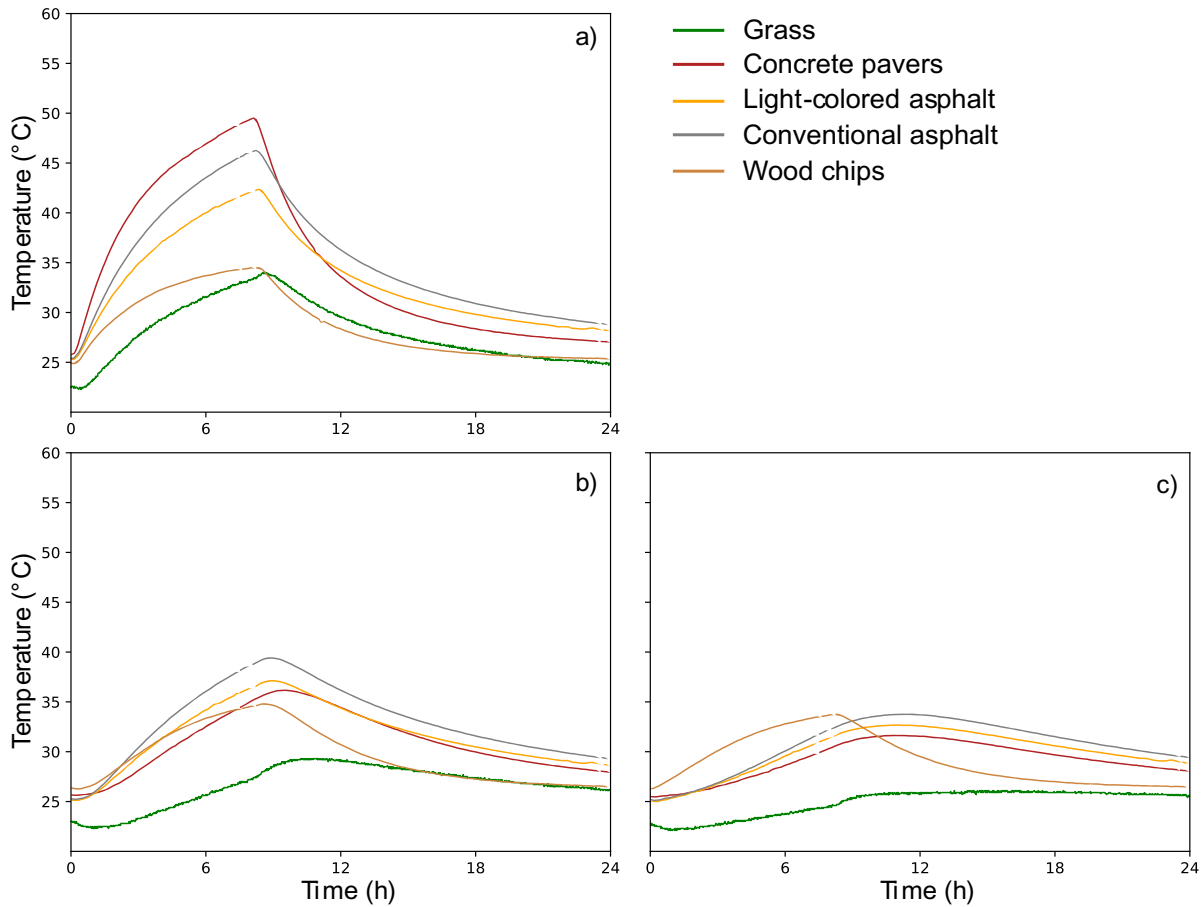

Figure 6: Temperature within samples at different depths: 6 cm (a), 14 cm (b) and 25 cm (c).

At a depth of 6 cm (Figure 7a), the temperatures of the wood chip structure and the lawn are cooler than those of the mineral materials. During the day, their temperatures are around 35°C maximum, dropping to around 25°C in the night phase, whereas concrete pavers, conventional asphalt and light-colored asphalt reach 50°C, 45°C and 40°C respectively in the day phase, before dropping back to temperatures between 30°C and 27°C in the night phase. The porosity of the soil and wood chips may partly explain this result, notably due to their lower thermal conductivity (as well as poorer thermal contact with the sensor). It is interesting to note that concrete pavers have a higher temperature at -6 cm than conventional asphalt. This result is attributable to the structure of concrete pavers, which appears to offer better conductivity than asphalt.

At a depth of 14 and 25 cm (Figure 7b and 7c), maximum temperatures within the wood chips are around 34°C 14 and 25 cm deep, confirming their low thermal conductivity.

The temperature distribution within the wood chip sample and the near absence of a phase shift between -6 cm and -25 cm leads us to question the effectiveness of the sample's lateral insulation. Indeed, the high thermal resistance of the chips could encourage lateral heat leakage through the thin insulating layer which may offer less thermal resistance. In addition, convection inside the sample pores may also explain part of the observed homogeneous temperatures, although during the day the top of the sample is thermally stratified and the pores are not very large given that the wood chips are 5-10 mm thick and a few centimeters long.

The temperature within the grass sample is also very attenuated at depths of 14 and 25 cm. The structure of the specimens explains the shape of the temperature curves within the mineral samples, with similar dynamics and temperature values for concrete pavers and light-colored asphalt. The temperature profile of conventional asphalt is similar to that of the other two mineral samples, but its maximum temperature is around 39°C, compared with around 36°C for the pavers and light-colored asphalt. The order of temperatures between the asphalt sample and the concrete pavers was again reversed, due to the porous sand layer between the concrete pavers and the soil subgrade, which has a lower thermal conductivity than the asphalt's concrete layer. The temperature curves of the mineral materials show the same pattern at -25 cm, with deviations of around 1.5°C at the end of the night phase.

**Heat Flux**

Heat flow within the samples is observed 6 and 14 cm deep, with the exception of the grass sample. These parameters are shown in Figure 8. Overall, the measured heat flows show similar patterns between samples, with a charge and discharge cycle corresponding to the diurnal and nocturnal phases of the experiment.

For mineral pavements, heat flows are greater at a depth of 6 cm for the concrete pavers than for the conventional asphalt, following the temperature differences, themselves linked to differences in albedo and thermal properties. During the day, the average flow is in the order of 80 $W/m^2$ for the concrete pavers, compared with 65 and 55 $W/m^2$ for conventional and light-colored asphalt, respectively.

Heat flow within the concrete paver sample is significantly lower 14 cm deep than 6 cm deep, with a maximum value of 20 $W/m^2$. This is attributed to the lower conductivity of the sand and gravel layer found beneath the pavers at this depth. Conversely, the concrete foundation of both asphalt structures reaches a steady state, as shown by the heat flows 6 and 14 cm deep equalizing by the end of the day phase.

Once again, the wood chips' low conductivity explains the low heat flows observed in that sample. The flux 6 cm deep is in the order of 10 $W/m^2$ and nearly zero 14 cm deep.

The heat flux observations support the conclusions drawn from the temperature distribution observed within the wood chips sample. In particular, their low conductivity explains the marked difference between surface temperature and temperature 6 cm deep. The wood chips sample was dry when subjected to the measurement cycle. Given the medium, a change in humidity would likely have an impact on the observed temperatures and heat flows. Consequently, summer rainfall even a few days beforehand or watering from fountains or children's water games may affect the structure's thermal performance during a heatwave.

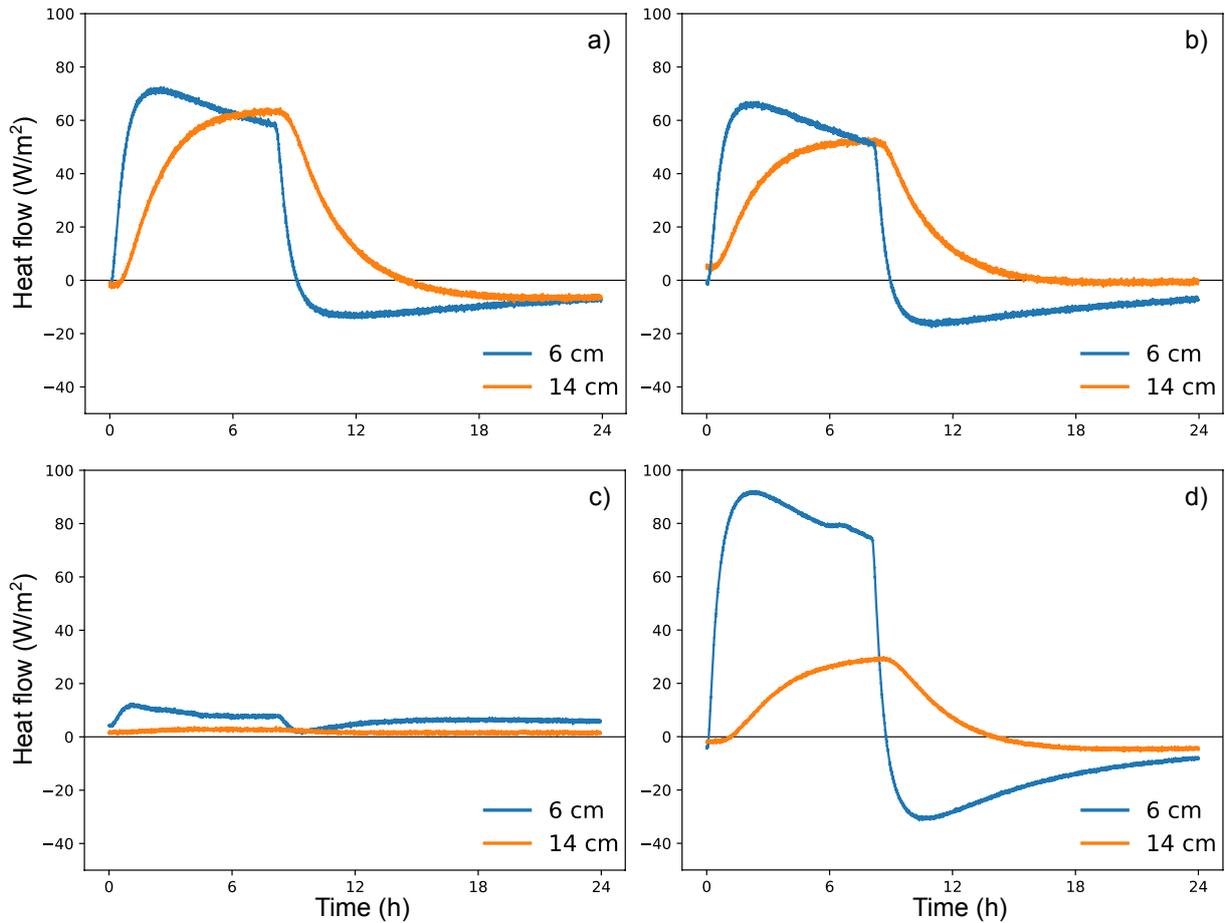

Figure 7: Heat flux density 6 cm (blue) and 14 cm (orange) deep for the conventional asphalt (a), light-colored asphalt (b), wood chips (c) and concrete pavers (d).

The transposition of these results to the field depends on conditions under which the material is deployed and used *in situ*. In schoolyards, wood chips are partly mixed with soil and compacts with use. This compaction and ageing is expected to contribute to an increase in thermal conductivity of the wood chips, but also to a reduction in surface albedo. Furthermore, the material is biodegradable, with a dense, moist compost layer 1 to 2 cm thick and 6 to 7 cm deep below the surface.

## Microclimatic Contributions

By observing temperatures and heat flows within the pavement structures, we can determine the terms of the surface heat balance, in particular radiosity and convection. By comparing convective exchanges and radiosity, it is possible to evaluate the materials' microclimatic contributions and suitability to mitigate UHI or improved pedestrian heat stress.

### Convective Contribution

Figure 9 summarizes the convective contributions observed over an experimental cycle for the various materials for daytime (red) and nighttime (blue), normalized by their irradiance.

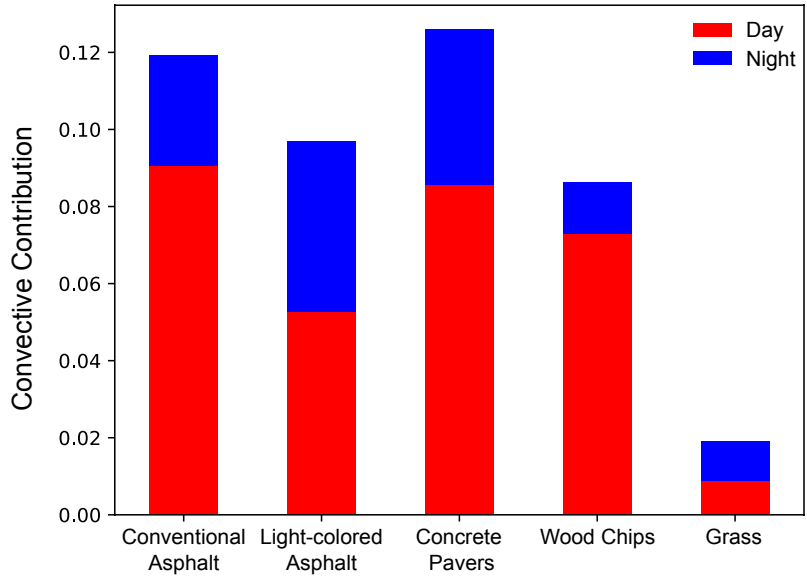

Figure 8: Nighttime (blue) and daytime (red) convective contributions

The convective contribution is calculated from convection *H*, determined according to Equation 5, and total irradiance *S+L*. It corresponds to cumulative convection over the duration of the experiment normalized by the cumulative total irradiance. Over 24 hours, concrete pavers and conventional asphalt contribute most to atmospheric heating, with around 12% of irradiance returned to the ambient air via convection. Wood chips and light-colored asphalt have an intermediate contribution, at 9% and 10% respectively. Finally, grass has the lowest contribution, with a convective contribution of less than 2% of its irradiance, divided equally between daytime and nighttime. This low value is due to the evapotranspiration of the grass, which evacuates a significant proportion of the absorbed heat, in the order of 10% given its radiative contribution (see below).

During the day, conventional asphalt and concrete pavers have very similar convective contributions in the order of 9% of their irradiance, closely followed by wood chips (7%) and light-colored asphalt (5%). At night, the contribution of wood chips is as low as that of grass, which can be explained by its low heat storage capacity, as observed in Figure 8. Light-colored asphalt and concrete pavers have the highest night-time convective contribution, due to the large amount of heat stored during the day, as illustrated by the heat flux measurements in Figure 8.

The higher night-time convection of light-colored asphalt compared with conventional asphalt is surprising at first glance considering its higher albedo. This could be due to a difference in the thermal properties of the light-colored binder, or a better thermal contact between the light-colored asphalt layer and its concrete mid-course, resulting in a higher thermal admittance for this sample.

**Radiative Contribution**

Figure 10 summarizes the values of radiosity *J* accumulated over an experimental cycle and normalized by the incident radiation $S + L$ (see Equation 7). Day (red) and night (blue) contributions as well as short- (orange) and long-wave radiosity are identified with different colors.

Firstly, the radiative contributions of the various samples are significantly higher than their convective contributions, ranging from 0.72 to 0.94 for the former, compared with values below 0.13 for the latter.

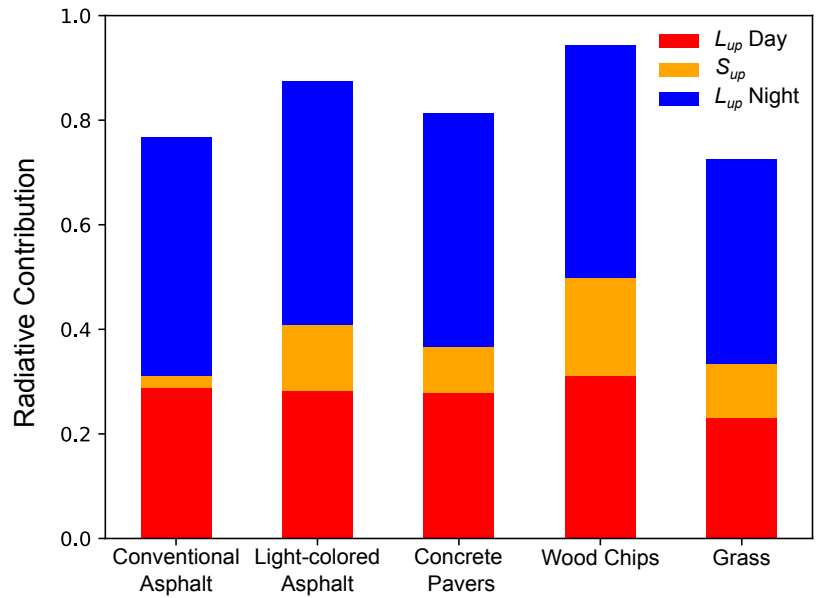

Figure 9: Cumulative *SW* and *LW* daytime (red or orange) and night-time (blue) normalized radiance

Most of the incident radiation is therefore returned to the environment via pavement radiosity: either infrared at long wavelengths (*LW*), or reflected solar radiation at short wavelengths (*SW*). Emitted infrared radiation accounts for the overwhelming majority of cumulative radiosity for all samples, ranging from 80% of total radiosity for the wood chips (with albedo 0.58) to 97% for the conventional asphalt.

During the day, the radiative contributions of the different samples range from 30% for conventional asphalt to 50% for wood chips, most of these differences being caused by differences in albedo which directly affect $S_{up}$, varying from 2 to 18%.

The relative contribution to $S_{up}$ of the wood chips sample is significantly higher than that of the other materials. This is due to its high albedo as measured in the laboratory. Field albedo measurements for wood chips indicate an albedo in the order of 0.25, which would approximately halve its *SW* radiosity. The wood chips would then have a diurnal *SW* radiosity of 8% rather than 19%, reducing its overall diurnal radiosity to 39% rather than 50%, i.e. a total radiative contribution drop from 0.94 to 0.84. The latter value is more consistent with its cumulative normalized convection (0.09), as the sum of the two terms cannot exceed a value of 1, as per Equation 10.

*LW* ($L_{up}$) radiosity ranges from 23% to 31% during the day and from 39% to 46% at night across all samples. This small variation is due to the small differences in absolute surface temperatures, raised to the power of 4, between the studied samples. Grass has the lowest radiosity of all the pavements studied, both day and night.

**Contributions to UHI Mitigation and Pedestrian Heat Stress**

These microclimatic indicators provide information on the microclimatic performance of the studied materials. In terms of UHI mitigation, grass, followed by wood chips and light-colored asphalt contribute the least to atmospheric heating over 24 hours, while conventional asphalt and concrete pavers contribute the most. Grass also offers low radiosity day and night, making it the best choice from a microclimatic point of view.

However, its generalized deployment in Parisian schoolyards is limited by the high intensity of their use by children during recess, while its thermo-climatic performance relies on the high availability of water for evapotranspiration. Water-stressed grass loses its ability to evapotranspire and its thermal performance is similar to that of the wood chips structure. Previous experiments with an unwatered grass sample revealed very high surface temperatures (58°C) and significantly lower temperatures below the surface [13].

The behavior of other the structures in terms of UHI (atmospheric heating) and heat stress (radiosity) is not aligned, day or night.

From this point of view, conventional pavements (conventional asphalt and concrete pavers) cause high atmospheric heating, both day and night, but have relatively low radiosity, comparable to that of the grass during the day. Conventional asphalt has the lowest daytime radiosity of all the samples studied (31%).

The other pavements (light-colored asphalt and wood chips) exhibit the opposite behavior (low atmospheric heating and high radiosity), with a significant contrast between day and night. Thus, while light-colored asphalt offers a low convective contribution during the day, it is significant at night. Furthermore, we can see that the benefit of light-colored asphalt in terms of atmospheric heating is obtained in exchange for greater radiosity. The same applies to wood chips, which combine high surface temperature and high radiosity in full sunlight, even if these terms are reduced at night. This high radiosity has a negative impact on pedestrian heat stress, particularly during the day when pedestrians are in the sun.

The contradictory effects of reduced atmospheric heating and increased radiosity need to be monitored very carefully before generalizing the use of pavements with such divergent behavior. Determining the total impact of these opposing trends on a pedestrian's heat stress level would require a numerical simulation with a more precise contextualization of the pedestrian and the pavement, particularly in terms of the pedestrian's position, the site's characteristics and morphology, the building façade materials, presence of shading devices, etc. However, a growing body of literature as well as mobile microclimatic measurements suggest that the increase in albedo is detrimental to pedestrian heat stress during the day, particularly for light-colored asphalt [18], [24].

From an operational perspective, the use of light-colored asphalt also needs to be validated, insofar as the lowering of its surface temperature is the result of its high albedo and therefore to its prolonged maintenance. Indeed, ageing and soiling tend to reduce its albedo [14], [25], and thus the benefits in terms of surface temperature. Wood chips seem to offer a more practical alternative from this point of view, but they come with their own maintenance constraints. Indeed, the manufacturers recommend that they should not be placed in full shade to slow their biological degradation and that new chips should be added periodically to compensate for their compaction. Therefore, placing them in the shade to limit their diurnal microclimatic impact (high convective contribution) is not compatible with this recommendation.

## Conclusion

A study of the thermal behavior of OASIS pilot schoolyard pavements under heatwave-like temperature and humidity conditions was conducted. Five pavement structures were tested: conventional asphalt, light-colored asphalt, concrete pavers, wood chips and water-saturated grass. The samples were instrumented with surface and in-depth temperature and heat flux sensors. Spectrophotometer and field albedo measurements were used to characterize their surface properties.

The coolest material is undoubtedly water-saturated grass, thanks to the dissipation of latent heat by evapotranspiration. Compared with other materials, wood chips have a characteristically short heating and cooling time. While they very quickly reach their highest surface temperature during the day (55°C), they cool down the quickest and show little heat accumulation. Light-colored asphalt achieves lower surface temperatures than conventional asphalt and has lower diurnal heat fluxes. On the other hand, due to its high albedo, light-colored asphalt has high radiosity during the day. Lastly, concrete pavers have a significant contribution to atmospheric heating during the day and the night, with high surface temperatures during the day and high heat storage, as evidenced by its high in-depth temperatures.

Laboratory studies of the thermal behavior of paving materials can be used to fine-tune pavement material selection according to the behavior best suited to the anticipated use periods of the target space and thus prioritizing pedestrian heat stress or UHI mitigation.

If the aim is to improve thermal comfort for users during the day in a sun-exposed area, this requires reducing the material's radiosity first and foremost, in addition to its surface temperature. The laboratory study shows that the grass is the most effective material for reducing thermal stress. However, its mechanical strength is limited and may not be compatible with heavily-used areas of Parisian schoolyards. High-albedo surfaces, on the other hand, risk increasing the average radiant temperature of users and may cause glare, likely compensating the benefits of their lower surface temperature, as seen in the field with light-colored asphalt. From an operational point of view, the problems of ageing and soiling of reflective pavements also question the relevance of these materials. In terms of pedestrian heat stress, concrete pavers perform less well than conventional asphalt and store heat significantly.

If mitigating the urban heat island effect is the priority, materials with low inertia and/or high albedo, which minimize daytime heat storage, are more appropriate. These include grass, wood chips and, to a lesser extent, light-colored asphalt. However, the effectiveness of high-albedo pavement assumes that their albedo remains stable over time and requires a high sky view factor. Indeed, the aim is to reflect solar irradiance towards the sky without interception by an obstacle such as a building, tree or shading device. Depending on the type of obstacle, the reflection may provide outdoor gains but aggravate building heat load, increase urban tree stress or simply be absorbed onsite, thus providing smaller-than-anticipated microclimatic gains for pedestrians.

In this article, mineral pavement materials were studied under dry conditions. However, surface temperatures can be lowered by prior humidification or watering, thanks to water evaporation [13]. Water availability is also an important factor in the study of vegetated or permeable samples. Analyzing the dry, wet or water-saturated behavior of such structures is essential to understanding their cooling performance.

The study of irrigated or watered materials would complete the analysis of the thermal behavior of the materials in order to optimize their implementation in future OASIS schoolyards. This would also enlighten local authorities as to the most suitable cooling techniques for schoolyards, taking into account usage constraints. A study of the impact of water on OASIS pavement materials would, for example, make it possible to account for the cooling effect of rainwater retained after a summer rainstorm or the benefits of water games and sustainable drainage systems. For instance, the Emeriau OASIS preschool schoolyard pictured in Figure 1 includes an educational artificial stream which the children can play with autonomously. Its use may provide indirect cooling benefits from watering low vegetation or the pavement materials.

Furthermore, the cumulative effects of repeated heatwave days has not been studied in this analysis. While the lab experiment could be repeated over several days, the realism of the experimental conditions is limited, in particular night time radiative cooling. An outdoor test site has been recently instrumented in collaboration with Paris City Hall and will help shed light on these aspects.

# Acknowledgements

The authors would like to acknowledge the significant support of the Roads and Traffic Division of the City of Paris for their expertise and material support. Funding for this study was provided by the ERDF UIA-0344-OASIS project.